# Synthesis of antisymmetric spin exchange interaction and entanglement generation with chiral spin states in a superconducting circuit

**Authors:** Da-Wei Wang[1,2,7,*], Chao Song[1], Wei Feng[1,3], Han Cai[1,2], Da Xu[1], Hui Deng[4,5], Dongning Zheng[4,5], Xiaobo Zhu[6,†], H. Wang[1,‡], Shiyao Zhu[1,6], and Marlan O. Scully[2]

**Affiliations:**

[1]Department of Physics, Zhejiang University, Hangzhou, Zhejiang 310027, China.

[2]Texas A&M University, College Station, Texas 77843, USA.

[3]Beijing Computational Science Research Center, Beijing 100193, China.

[4]Institute of Physics, Chinese Academy of Sciences, Beijing 100190, China.

[5]School of Physical Sciences, University of Chinese Academy of Sciences, Beijing 100049, China.

[6]Synergetic Innovation Center of Quantum Information and Quantum Physics, University of Science and Technology of China, Hefei, Anhui 230026, China.

[7]CAS Center of Excellence in Topological Quantum Computation, Beijing 100190, China.

Correspondence to:   *dwwang@zju.edu.cn, †xbzhu16@ustc.edu.cn, ‡hhwang@zju.edu.cn

**Abstract**: We have synthesized the anti-symmetric spin exchange interaction (ASI), which is also called the Dzyaloshinskii-Moriya interaction, in a superconducting circuit containing five superconducting qubits connected to a bus resonator, by periodically modulating the transition frequencies of the qubits with different modulation phases. This allows us to show the chiral spin dynamics in three-, four- and five-spin clusters. We also demonstrate a three-spin chiral logic gate and entangle up to five qubits in Greenberger-Horne-Zeilinger states. Our results pave the way for quantum simulation of magnetism with ASI and quantum computation with chiral spin states.

**Main Text:** In order to explain the rotation of the polarization plane of light in crystals, Louis Pasteur introduced molecular chirality which differentiates two molecular structures that are non-superposable mirror images of each other. According to quantum mechanics, chiral states cannot be stationary states of a parity conserving molecular Hamiltonian. This is in contradiction to the existence of optical isomers, as questioned by Hund [1]. The origin of molecular and biological chirality and its relation to fundamental parity violation and environmental decoherence is still under debate [2,3]. A quantum superposition of two chiral molecular states with distinctive properties such as their optical activity has never been observed [4]. The difficulty lies in how to implement a controllable potential barrier between two physically discernable chiral states. For molecules with a small potential barrier such as ammonia, the ground and first excited states are symmetric and anti-symmetric superpositions of the two chiral states. The transition between these two eigenstates was used in the first maser [5]. However, the two chiral states cannot be directly measured. For optical isomers such as tartaric acid, the high potential barrier and complex molecular structure render a quantum superposition of the two chiral states literally impossible.

Two conditions are needed for the experimental observation of a quantum superposition of different chiral states. First, the chiral states must have observables that characterize their chirality. Second, there must be controllable interactions between the two chiral states to prepare a quantum superposition. A suitable platform to satisfy these conditions is the superconducting circuit, which has been widely used in quantum simulation and quantum computation with advantageous tunability, flexibility, and scalability [6,7]. In particular, chiral ground-state currents of interacting photons hosted by three qubits were observed with a synthesized magnetic field [8]. The ground state energy of molecules containing three atoms has been calculated in a superconducting quantum processor [9]. However, quantum logic operations based on chiral spin states have never been demonstrated despite their being of substantial theoretical interest [10,11].

Recently we have shown [12] that if we initially prepare an unentangled state $|\psi\rangle = |N,0,0\rangle(|e\rangle+|g\rangle)/\sqrt{2}$ where $|N,0,0\rangle$ is a three-cavity Fock state, and $|g\rangle$ and $|e\rangle$ are

the ground and excited atomic states (pseudo-spin states), we can synthesize a Hamiltonian with spin-gated chirality that results in an mesoscopic entangled state $|\psi\rangle = (|0, N, 0\rangle |e\rangle + |0, 0, N\rangle |g\rangle)/\sqrt{2}$. Following along similar lines we have experimentally demonstrated parallel logic operations with superpositions of chiral states in a superconducting circuit containing 5 qubits. The observable chiral behavior is realized based on a synthesized Dzyaloshinskii-Moriya interaction [13, 14], which results in chiral spin currents with spin-dependent chirality. Dzyaloshinskii-Moriya interaction is the antisymmetric spin exchange interaction (ASI) which plays an important role in establishing exotic spin orders and is currently under intensive experimental investigation in magnetic materials [15-21]. In contrast to the symmetric Heisenberg interaction, which has been realized years ago [22] and recently used to entangle 10 superconducting qubits [23], the synthesis of an ASI has never been realized, although a photonic gauge field with photon-photon interactions has a similar effect [8]. Here we synthesize an ASI by periodically modulating the transition frequencies of three qubits coupled to the same bus resonator. The ASI Hamiltonian violates parity symmetry but conserves time-reversal symmetry. The parity symmetry breaking ensures chiral spin currents, while the conservation of time-reversal symmetry guarantees opposite chiral dynamics of two spin configurations. This property is in analogue to the quantum spin Hall effect [24], where electrons with opposite spins travel in opposite directions on the edges of a topological insulator. We have observed chiral spin dynamics in spin-clusters containing up to 5 spins. By preparing a superposition of spin states with opposite chiral dynamics, we first entangle 3 qubits in GHZ states and then demonstrate its scalability to 5 qubits.

To motivate the underlining physics, we first introduce a central concept in describing chiral spin configurations, the spin chirality [25-28],

$$C_z = \frac{1}{2\sqrt{3}} \boldsymbol{\sigma}_1 \cdot (\boldsymbol{\sigma}_2 \times \boldsymbol{\sigma}_3), \qquad (1)$$

where $\boldsymbol{\sigma}_j = \sigma_j^x \hat{x} + \sigma_j^y \hat{y} + \sigma_j^z \hat{z}$ ($j=1,2,3$) is the Pauli vector for the $j$th spin. Classically this quantity is proportional to the solid angle subtended by the three spins [26]. With chirality as a Hamiltonian, the three classical spins precess around their central axis. In quantum mechanics, the dynamics follows a similar behavior as shown by the exponential operator of $C_z$, which permutes the three spin states in a chiral way, i.e.,

$$e^{-iC_z\theta/2} |s_1 s_2 s_3\rangle = |s_3 s_1 s_2\rangle, \tag{2}$$

where $s_j = \uparrow$ or $\downarrow$ and $\theta = 4\pi/3$. This operation is nontrivial only when one of the spin states is different from the other two, such as $|\uparrow\downarrow\downarrow\rangle$ and $|\downarrow\uparrow\uparrow\rangle$.

The chirality operator $C_z$ breaks both parity and time reversal symmetry. In order to observe opposite chiral dynamics of the two configurations $|\uparrow\downarrow\downarrow\rangle$ and $|\downarrow\uparrow\uparrow\rangle$, we need a Hamiltonian that breaks parity symmetry but conserves time reversal symmetry. The product of $C_z$ and $S_z = \sum_{j=1}^{3} \sigma_j^z / 2$ satisfies this requirement,

$$H = \hbar \kappa S_z C_z = \sum_{j=1}^{3} \mathbf{D} \cdot (\boldsymbol{\sigma}_j \times \boldsymbol{\sigma}_{j+1}), \tag{3}$$

where $\mathbf{D} = \hbar \kappa \hat{z}/4\sqrt{3}$ with $\kappa$ being a coupling constant. Eq. (3) is a chiral ASI Hamiltonian of three spins in a configuration schematically shown in Fig.1 (A). The key feature of $H$ is that the two subspaces with $S_z = \pm 1/2$ have opposite chiral dynamics,

$$\begin{aligned} e^{-iHT_0/\hbar} |\uparrow\downarrow\downarrow\rangle &= |\downarrow\downarrow\uparrow\rangle, \\ e^{-iHT_0/\hbar} |\downarrow\uparrow\uparrow\rangle &= |\uparrow\downarrow\uparrow\rangle, \end{aligned} \tag{4}$$

where $T_0 = \theta/\kappa$. To understand this, we note that the two opposite chiral evolutions are time-reversal to each other.

The experimental results of the dynamics in Eq. (4) are shown in Fig. 2 with a chip containing 5 transmon qubits interconnected by a central bus resonator. We periodically modulate the transition frequencies of the three qubits in focus ($Q_1$, $Q_2$ and $Q_4$ in Fig. 4) and set the other two qubits far detuned from the resonator. The time-dependent transition frequencies of the three qubits are $\omega_j(t) = \omega_0 + \Delta\cos(\nu t - \phi_j + \phi_0)$ with $\Delta/2\pi = 235$ MHz, $\nu/2\pi = 98.8$ MHz, $\phi_j = 2j\pi/3$, and $\phi_0$ being an initial phase, which results in an effective Floquet Hamiltonian $H$ in Eq. (3) with $\kappa = 2\pi \times 4.44$ MHz and $T_0 = 150$ ns. A calculation of the effective Floquet Hamiltonian [29, 30] and the experimental details are in the Supplementary Materials.

The chiral dynamics in Fig. 2 can be explicitly calculated from the dispersion relation of the Hamiltonian, as shown in Fig. 1 (B). In the subspace of $S_z = \pm 1/2$, the eigenenergies are $\lambda^\pm(k) = \pm k\hbar\kappa/2$ with $k = 0, \pm 1$ being the eigenvalues of $C_z$ and the corresponding spin-wave eigenstates are $|k, -1/2\rangle = (|\uparrow\downarrow\downarrow\rangle + e^{2ik\pi/3}|\downarrow\uparrow\downarrow\rangle + e^{4ik\pi/3}|\downarrow\downarrow\uparrow\rangle)/\sqrt{3}$ and $|k, +1/2\rangle = (|\downarrow\uparrow\uparrow\rangle + e^{2ik\pi/3}|\uparrow\downarrow\uparrow\rangle + e^{4ik\pi/3}|\uparrow\uparrow\downarrow\rangle)/\sqrt{3}$. By uniformly placing the three spins on a unit circle and labelling their positions by $0, 2\pi/3, 4\pi/3$, we can understand $k = 0, \pm 1$ as the momenta of the spin waves. In considering the dynamics of a spin configuration $|\uparrow\downarrow\downarrow\rangle$ or $|\downarrow\uparrow\uparrow\rangle$, which are spin wave packets, their group velocities are,

$$v_g^\pm = \frac{\partial \lambda^\pm(k)}{\hbar \partial k} = \pm\kappa/2. \qquad (5)$$

It costs a time $T_0 = 2\pi/3|v_g^\pm| = 4\pi/3\kappa$ for the two spin configurations to move for one step (spin site) in opposite directions, which is consistent with Eq. (4). This analysis is in reminiscence of the quantum spin Hall effect [24]. The three spins in Fig. 1 (A) can be viewed as a minuscule two-dimensional lattice with only edges. The three states in each subspace $S_z = \pm 1/2$ are chiral edge states of spin waves that host edge currents in opposite directions.

The dynamics in Fig. 2 can also be interpreted as a precession of the chirality vector [28], $\mathbf{C} = C_x \hat{x} + C_y \hat{y} + C_z \hat{z}$ where $C_x = (2\boldsymbol{\sigma}_2 \cdot \boldsymbol{\sigma}_3 - \boldsymbol{\sigma}_1 \cdot \boldsymbol{\sigma}_2 - \boldsymbol{\sigma}_3 \cdot \boldsymbol{\sigma}_1)/12$ and $C_y = (\boldsymbol{\sigma}_1 \cdot \boldsymbol{\sigma}_2 - \boldsymbol{\sigma}_3 \cdot \boldsymbol{\sigma}_1)/4\sqrt{3}$. The three components have the same commutation relation as the Pauli matrices, $[C_i, C_j] = 2\varepsilon_{ijk} C_k$. Therefore, the Hamiltonian $H$ can be understood as that $\mathbf{C}$ is subjected to a field along $\hat{z}$ axis with directions depending on $S_z$. The six states $|k, S_z\rangle$ can be grouped into two singlets and two doublets of the eigenstates of $\mathbf{C}$. We initially prepare the spins in $|\uparrow\downarrow\downarrow\rangle \equiv |0\rangle/\sqrt{3} + \sqrt{2}|\Uparrow_x\rangle/\sqrt{3}$, where $|0\rangle \equiv |0, -1/2\rangle$ is a singlet eigenstate and does not evolve with time, while $|\Uparrow_x\rangle = (|1, -1/2\rangle + |-1, -1/2\rangle)/\sqrt{2}$ is a doublet eigenstate of $C_x$ with the eigenvalue $C_x = 1$ and precesses on the equator of the chirality Bloch sphere around $S_z \hat{z}$ with $S_z = -1/2$ (and thus the precession is in the opposite direction in the $S_z = 1/2$ subspace), as shown in Fig.2. After every rotation of $\theta = 4\pi/3$, the wavefunction evolves to a separable state $|\downarrow\downarrow\uparrow\rangle$ and then $|\downarrow\uparrow\downarrow\rangle$ in a chiral way.

The chiral dynamics of spin clusters containing 4 and 5 spins with ASI is shown in Fig. 3. Substantially different from the symmetric interaction where a spin excitation undergoes Bloch oscillations with a superposition (bright) state of other spins, the ASI lifts the degeneracies of parity-symmetric eigenstates and allows a chiral evolution of the spin excitation over all spin sites in a sequence determined by the order of $i$ and $j$ in $\boldsymbol{\sigma}_i \times \boldsymbol{\sigma}_j$. In Fig. 3 (A), the interaction Hamiltonian is $H = \mathbf{D} \cdot (\boldsymbol{\sigma}_2 \times \boldsymbol{\sigma}_1 + \boldsymbol{\sigma}_4 \times \boldsymbol{\sigma}_1 + \boldsymbol{\sigma}_1 \times \boldsymbol{\sigma}_3 + \boldsymbol{\sigma}_3 \times \boldsymbol{\sigma}_2 + \boldsymbol{\sigma}_3 \times \boldsymbol{\sigma}_4)$, which can be achieved by setting the modulation phases $\phi_1 = 0$, $\phi_2 = \phi_4 = -2\pi/3$ and $\phi_3 = 2\pi/3$. If we initially prepare spin 1 in the spin-up state and all other spins in spin-down states, the spin-up excitation chirally evolves to a superposition of spins 2 and 4, then to spin 3, and finally returns to spin 1. In each of the triangle, there is chiral evolution similar to Fig. 2. If we reverse the initial state of all spins, the chiral evolution has an opposite direction due to the time-reversal symmetry, as shown in Fig. 3 (B). Although three spin-up states have a tripled decay rate, the chiral evolution is still obvious. In Fig. 3 (C), we have a five-spin cluster, where a central spin 1 interacts with four other spins with the ASI between the nearest neighbors. The Hamiltonian is $H = \mathbf{D} \cdot (\boldsymbol{\sigma}_2 \times \boldsymbol{\sigma}_1 + \boldsymbol{\sigma}_4 \times \boldsymbol{\sigma}_1 + \boldsymbol{\sigma}_1 \times \boldsymbol{\sigma}_3 + \boldsymbol{\sigma}_1 \times \boldsymbol{\sigma}_5 + \boldsymbol{\sigma}_3 \times \boldsymbol{\sigma}_2 + \boldsymbol{\sigma}_5 \times \boldsymbol{\sigma}_2 + \boldsymbol{\sigma}_3 \times \boldsymbol{\sigma}_4 + \boldsymbol{\sigma}_5 \times \boldsymbol{\sigma}_4)$, which can be achieved by setting $\phi_1 = 0$, $\phi_2 = \phi_4 = -2\pi/3$ and $\phi_3 = \phi_5 = 2\pi/3$. In each of the triangle, there is still a chiral evolution determined by the interaction order in the ASI, such that the spin-up excitation first evolves from spin 1 to a superposition of spins 2 and 4, then to a superposition of spins 3 and 5, and finally comes back to spin 1.

A chiral rotation of photons in three cavities has been proposed [12, 31] and experimentally realized in superconducting qubits recently [8]. In particular, by modulating the coupling strengths between three cavities and taking advantage of the nonlinearity of photons, it has been shown that synthetic gauge field with interactions between photons results in different chiral dynamics for single photon and two-photon states [8], which is similar to the results in Fig. 2. However, the implementation of chiral spin states in quantum information has not been experimentally explored. In the following, we demonstrate the quantum parallel operation of entangling three qubits at a time and its scalability to five qubits.

The procedure of preparing a GHZ states and the experimental results are shown in Fig. 4. Starting with the state $|\downarrow\downarrow\downarrow\rangle$, we flip the first spin and apply a $\pi/2$ pulse to the second spin, which results in the wave function $|\Psi(0)\rangle = (|\uparrow\downarrow\downarrow\rangle + |\uparrow\uparrow\downarrow\rangle)/\sqrt{2}$. We then turn on the ASI Hamiltonian for the time $T_0$. Due to opposite chiral dynamics of the two component states, the wavefunction evolves to $|\Psi(T_0)\rangle = (|\downarrow\uparrow\downarrow\rangle + |\uparrow\downarrow\uparrow\rangle)/\sqrt{2}$. This step is a three-spin chiral logic gate $\text{ASI}(ijk)$ with $ijk$ denoting the spin sequence $123$ in $H$, as shown in Fig. 4.

Then we apply a $\pi$ pulse to the second qubit and obtain a 3-qubit GHZ state. The scalability of this scheme is demonstrated by adding two additional spins $|\uparrow\downarrow\rangle$ to the prepared 3-qubit GHZ state, $|\Psi(T_0)\rangle = (|\downarrow\downarrow\downarrow\uparrow\downarrow\rangle + |\uparrow\uparrow\uparrow\uparrow\downarrow\rangle)/\sqrt{2}$. Now we turn on the ASI Hamiltonian between the third, fourth and fifth spins for the time $T_0$, which results in the state $|\Psi(2T_0)\rangle = (|\downarrow\downarrow\uparrow\downarrow\downarrow\rangle + |\uparrow\uparrow\downarrow\uparrow\uparrow\rangle)/\sqrt{2}$. Then we apply a $\pi$ pulse to the third spin to prepare a 5-qubit GHZ state. The fidelities $F$ of the 3-qubit GHZ states prepared with spins $Q_1Q_2Q_4$ and $Q_2Q_3Q_5$ are $0.846\pm0.014$ and $0.877\pm0.012$, respectively. The fidelity of the 5-qubit GHZ state is $0.588\pm0.008$, which is above $50\%$ and satisfies the criteria for genuine entanglement [32]. The fidelity can be greatly improved with specifically designed circuit architectures.

Our experiment demonstrates a three-qubit chiral gate based on a synthesized ASI in a superconducting circuit. Complementing the symmetric interactions, the synthesis of ASI in a commonly used superconducting circuit paves the way for simulating chiral magnetism. The dynamics of such a system with a parity symmetry breaking Hamiltonian, in particular the decoherence of a superposition of different chiral states, can help us to understand the existence of chirality in general, e.g., chiral chemical and biological molecules. This work also provides a platform for conducting quantum computation with chiral spin states.

**Acknowledgments:** We thank Wuxin Liu, Qiujiang Guo and Keqiang Huang for technical support. This research was supported by the National Basic Research Program of China (Grants No. 2014CB921201 and 2014CB921401), the National Natural Science Foundations of China (Grants No. 11434008 , No. 11574380, and No. 11374344 ), and the Fundamental Research Funds for the Central Universities of China (Grant No. 2016XZZX002-01). S. Z. was supported by Joint Fund of National Natural Science Foundation of China (U1330203). D. W. was supported by the key research program of the Chinese Academy of Sciences (Grant No. XDPB08−3) and Robert A. Welch Foundation (Grant No. A-1261). H.C. was supported by the Herman F. Heep and Minnie Belle Heep Texas A&M University Endowed Fund held and administered by the Texas A&M Foundation. Devices were made at the Nanofabrication Facilities at Institute of Physics in Beijing, University of Science and Technology in Hefei, and National Center for Nanoscience and Technology in Beijing.

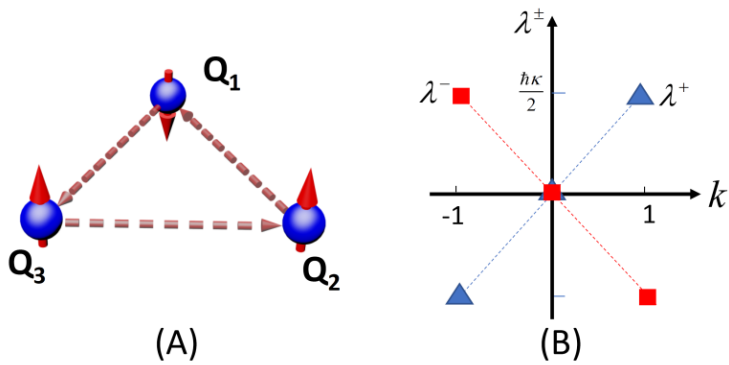

**Fig. 1**. **Spin configurations and spin wave dispersion relation.** (**A**) Three quantum spins $\mathbf{Q}_i$ ($i=1,2,3$) with chiral ASI (denoted by the dashed arrows. An arrow from $i$ to $j$ means the interaction $\boldsymbol{\sigma}_j \times \boldsymbol{\sigma}_i$), which can be viewed as either a one-dimensional tight-binding chain or a minuscule two-dimensional triangular ring. (**B**) The energy spectra of the Hamiltonian $H$ in Eq. (3). The blue triangles (red squares) are for $\lambda^+(k)$ and $\lambda^-(k)$, respectively.

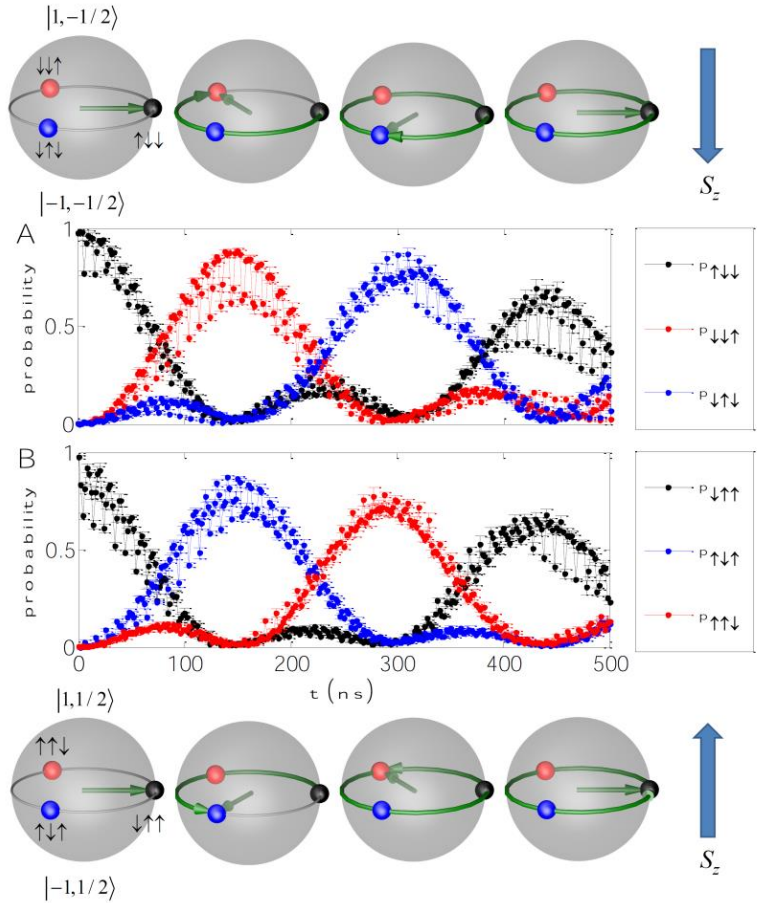

**Fig. 2. Experimental observation of the chiral spin dynamics induced by synthesized antisymmetric spin exchange interaction.** (**A**) Chiral dynamics with the three spins initialized in $|\uparrow\downarrow\downarrow\rangle$. The time-dependent probabilities of the spin configurations are shown as labeled. The one up spin moves chirally from $1 \rightarrow 3 \rightarrow 2$. (**B**) Chiral dynamics with the three spins initialized in $|\downarrow\uparrow\uparrow\rangle$. The one down spin moves from $1 \rightarrow 2 \rightarrow 3$, with an opposite chirality compared with

the case in **A**. The three spins 123 correspond to $Q_4 Q_2 Q_1$ in Fig. 3. The data were averaged over five tests. In the chirality Bloch spheres of the doublet states with $S_z = \pm 1/2$, we show how the chirality vector **C** (indicated by the green arrows) precesses around an effective field (in the direction of the thick blue arrows on the far right). The states labelled on the Bloch sphere are the resultant states from the interference between the doublet states and the singlet state, not the doublet states themselves. Please note that only after $4\pi$ rotation, the initial states are recovered, which is a property of the spin-1/2 nature of the doublet subspace. The total wave function at multiples of $T_0$ are shown by the small colored balls. The neighboring data are connected by straight lines to show the fast oscillations due to the modulation of the transition frequencies of the qubits. The numerical fitting and simulation can be found in the Supplementary Materials.

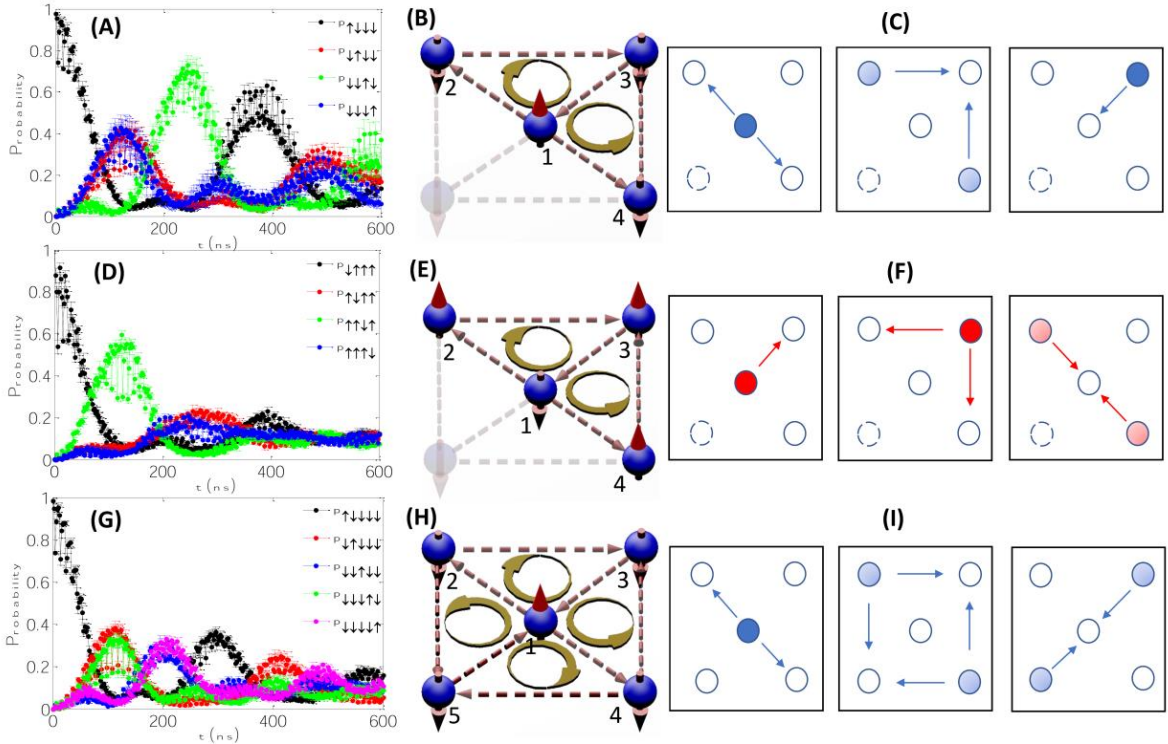

**Fig. 3. Chiral dynamics of spin excitation in 4- and 5- spin-clusters.** (**A**) The dynamic evolution of the initial state $|s_1 s_2 s_3 s_4\rangle = |\uparrow\downarrow\downarrow\downarrow\rangle$. (**B**) The initial spin configuration (green arrows) and the ASI between them (red arrows from $i$ to $j$ denote the interaction $\boldsymbol{\sigma}_j \times \boldsymbol{\sigma}_i$). The arrowed circles inside each triangle denote the chiral evolution direction of a single spin-up excitation. (**C**) Schematic chiral evolution of a single spin-up excitation. The depths of the blue color denote the probability of the spin-up states. The dashed circle denotes the fifth spin that does not interact with the others. (**D**)-(**F**) The evolution of a single spin-down state with opposite chiral dynamics compared with a single spin-up state. The depths of the red color in (**F**) denote

the probabilities of the spin-down states. **(G)-(H)** The chiral dynamics of a single spin-up state in a five-spin cluster with ASI.

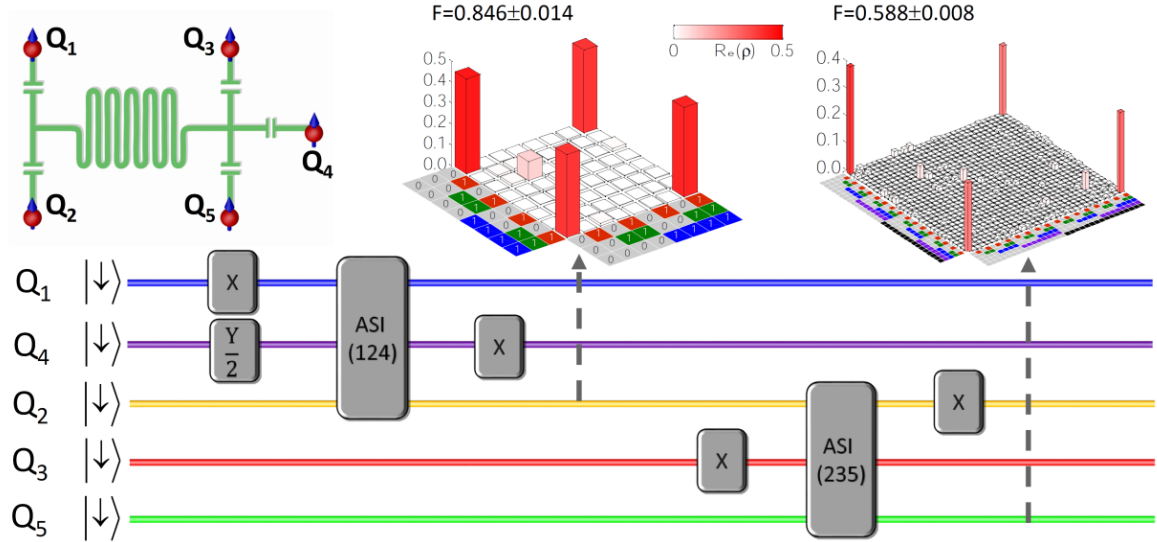

**Fig. 4. Quantum circuit and quantum state tomography for preparing Greenberger-Horne-Zeilinger states with parallel logic operations based on the antisymmetric spin exchange interaction.** Upper left is the circuit schematic of the five qubits (spins) connected with the bus resonator (central sinusoid lines). In the quantum circuits, $X$ is the Pauli-X gate (NOT) gate, $Y/2$ is the $\pi/2$ rotation gate around $\hat{y}$ axis, and ASI($ijk$) with $Q_i Q_j Q_k$ corresponding to spin 123 in Eq. (3) is the three-spin chiral gate. $F$ is the fidelity of the two GHZ states.

**Supplementary Materials:**

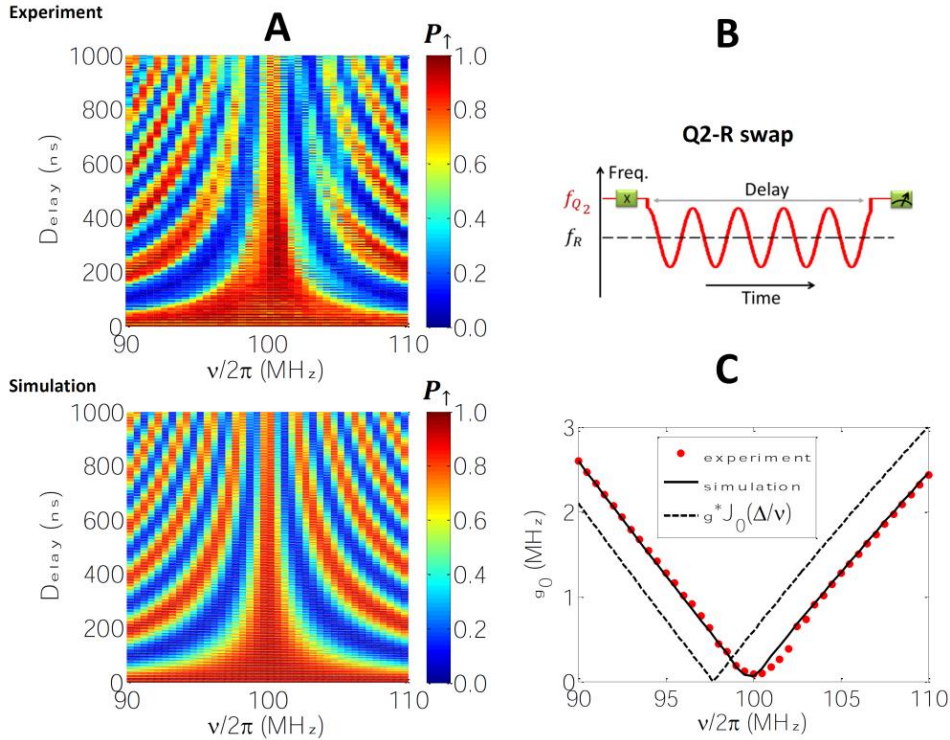

**Figure S1. Tunable effective coupling between a qubit and the resonator.** (**A**) Experimental data and simulation of the vacuum Rabi oscillations between the qubit $Q_2$ and the resonator with different modulation frequencies $\nu$. Shown are the qubit $|\uparrow\rangle$ state probabilities $P_\uparrow$ (see color bar on the right) as functions of the interaction (delay) time along y axis. (**B**) The pulse sequence for controlling the interaction between the qubit $Q_2$ and the resonator, with frequencies $f_{Q_2}$ and $f_R$, respectively. The box X is a X-gate for $Q_2$. (**C**) Effective coupling strength $g_0$ between the qubit and the resonator as a function of $\nu$ obtained by Fourier transform of the data in **A** (red dots). The dashed line is the analytic result with second order approximation, while the solid line is by numerical simulation that takes into account high order terms in the Hamiltonian. The modulation amplitude $\Delta/2\pi = 235$ MHz, and the decoupling point with $J_0(f) = 0$ is at $\nu/2\pi = 100$ MHz.

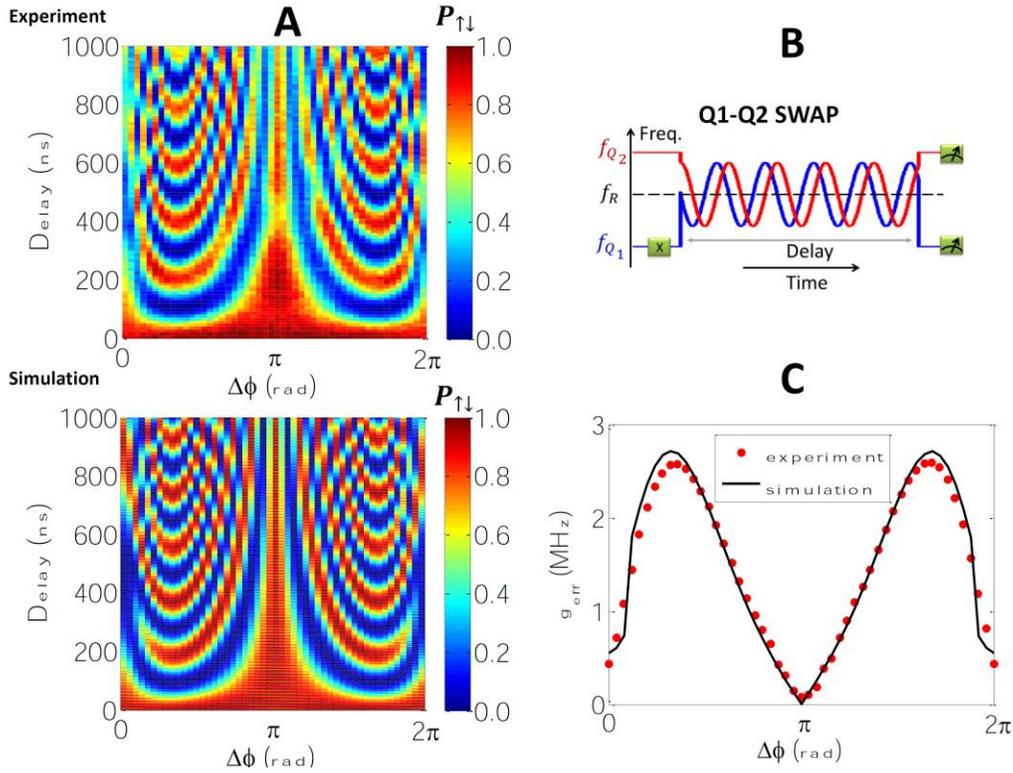

**Figure S2. Tunable effective coupling between two qubits.** (**A**) Experimental data and simulation of two-qubit swap dynamics as a function of the modulation phase difference between the two qubits, $\Delta\phi$. The probabilities that $Q_1$ is in the $|\uparrow\rangle$ state and $Q_2$ is in the $|\downarrow\rangle$ state, $P_{\uparrow\downarrow}$, are shown (see color bar on the right) as functions of the interaction (delay) time along y axis. (**B**) The pulse sequence that controls the interaction between the two qubits and the resonator. (**C**) Effective coupling strength as a function of the modulation phase difference, obtained by Fourier transform of the data in A. The modulation amplitude and frequency are $\Delta/2\pi = 235$ MHz and $\nu/2\pi = 100$ MHz, respectively.

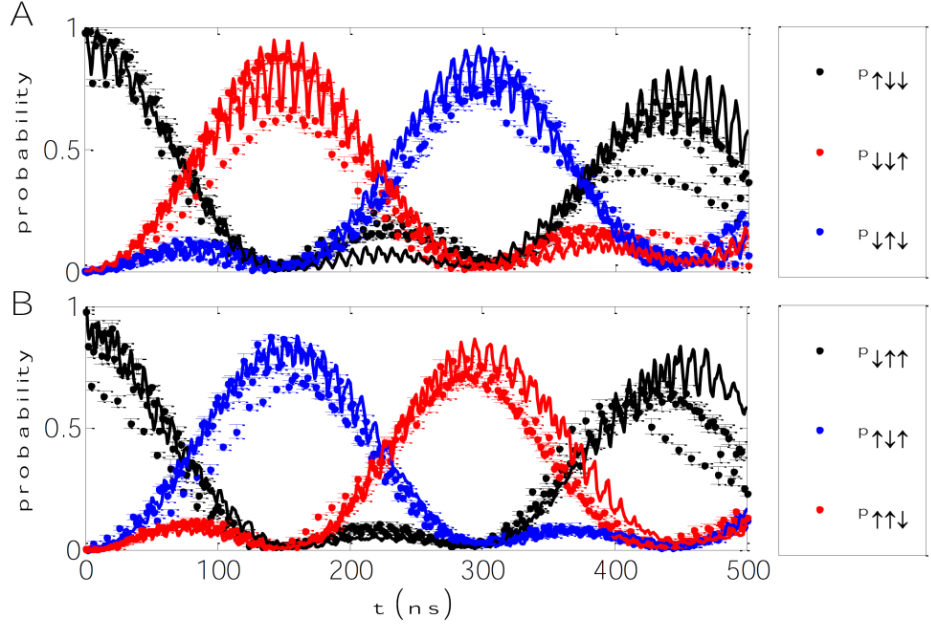

**Figure S3. Numerical fitting of the experimental data.** We have used the parameters in Tab. S1 and used fitting parameters for the coupling strength $g' = 1.03g$, qubit-cavity detuning $\delta = \omega_q - \omega_r = -3.5$ MHz and an overall initial modulation phase $\phi_0 = \pi/3$. The lines are the numerical results.

|  | $Q_1$ | $Q_2$ | $Q_3$ | $Q_4$ | $Q_5$ |
|---|---|---|---|---|---|
| $\omega_j^{idle}/2\pi$ (GHz) | 5.204 | 5.897 | 5.287 | 5.253 | 5.341 |
| $\eta_j/2\pi$ (MHz) | -245 | -242 | -245 | -243 | -244 |
| $g_j/2\pi$ (MHz) | 20.9 | 20.6 | 20.1 | 18.8 | 19.8 |

| $T_{1,j}$ (μs) | 20.2 | 10.2 | 18.9 | 19.2 | 13.9 |
| $T_{2,j}^{*}$ (μs) | 1.1 | 4.3 | 1.3 | 0.7 | 1.7 |

**Table S1**. Device parameters. $\omega_j^{idle}$ are the idle frequencies of the qubits, $\eta_j$ are the anharmonicities, $g_j$ are the qubit-resonator coupling strengths, $T_{1,j}$ are the qubit lifetimes, and $T_{2,j}^{*}$ are the Gaussian dephasing times of the qubits.

References:

[S1] J. Johansson, P. Nation, and F. Nori, Comput. Phys. Commun. 183, 1760–1772 (2012).

[S2] C. K. Law, S. Y. Zhu, and M. S. Zubairy, Phys. Rev. A 52, 4095-4098 (1995).

**Supplementary Materials:**

**Device** — Our sample contains 5 frequency-tunable superconducting transmon qubits interconnected by a bus resonator. The resonant frequency of the bus resonator is fixed at $\omega_r / 2\pi \approx 5.585$ GHz. The qubit transition frequencies are set at $\omega_j^{idle}$ (see Tab. S1) for state initialization and measurement, where single-qubit rotation gates are applied. The coupling strengths $g_j$ between the qubit and the bus resonator are listed in Tab. S1. The qubit anharmonicities are characterized by $\eta_j = \omega_j' - \omega_j^{idle}$, where $\omega_j'$ is the transition frequency between the first excited state and the second excited state. The resonator lifetime is $T_1^r \approx 13$ μs, and all qubit coherence parameters are listed in Tab. S1.

**Derivation of Eq. (3)** — The quantity

$$4\sqrt{3}S_zC_z = \sum_j \sigma_j^z \sum_{j'} \sigma_{j'}^z \left(\sigma_{j'+1}^x \sigma_{j'+2}^y - \sigma_{j'+1}^y \sigma_{j'+2}^x\right)$$

$$= \sum_j \left(\sigma_{j+1}^x \sigma_{j+2}^y - \sigma_{j+1}^y \sigma_{j+2}^x\right) + \sum_j \left(\sigma_j^z + \sigma_{j+1}^z\right)\left(\sigma_j^x \sigma_{j+1}^y - \sigma_j^y \sigma_{j+1}^x\right),$$

where we have used $\sigma_j^z \sigma_j^z = 1$ and the summation indices $j$ and $j'$ are cyclic from 1 to 3. The second summation is zero, $\left(\sigma_j^z + \sigma_{j+1}^z\right)\left(\sigma_j^x \sigma_{j+1}^y - \sigma_j^y \sigma_{j+1}^x\right) = 0$ since $\sigma_j^z \sigma_j^x = i\sigma_j^y$. The first term is $4\sqrt{3}\sum_{j=1}^{3} \mathbf{D} \cdot \left(\boldsymbol{\sigma}_j \times \boldsymbol{\sigma}_{j+1}\right)$, which proves Eq. (3).

**Effective Hamiltonian** — Due to the imperfections in the fabrication of the sample, the direct coupling strengths between the qubits and the bus resonator $g_j$ are slightly different for each qubit (as shown in Tab. S1). For the sake of simplicity, we approximate that they are all equal to $g = 20$ MHz. The interaction Hamiltonian of three qubits coupled with the same cavity under the rotating wave approximation is

$$H_I = \hbar g \sum_{j=1}^{3}\left(a^\dagger \sigma_j^- e^{-if\sin(vt - 2j\pi/3 + \phi_0)} + H.c.\right), \tag{S1}$$

where $a^\dagger$ and $a$ are the creation and annihilation operators of the resonator, and $f = \Delta/v$. The central frequencies of the qubits are equal to that of the resonator. Under the condition $v \gg g$, we obtain the effective Floquet Hamiltonian $H_I = H_0 + H$, where $H_0 = \hbar g J_0(f) \sum_{j=1}^{3}(a^\dagger \sigma_j^- + H.c.)$ and $H$ is defined in Eq. (3) with $\kappa = 2\sqrt{3}g^2\beta/v$ and $\beta = \sum_{n=1}^{\infty} 2J_n^2(f)\sin(2n\pi/3)/n$. Here $J_n(f)$ is the $n$th order Bessel function of the first kind. Ideally, when $f = 2.40$, $J_0(f) = 0$, we obtain $H_I = H$ with $\beta \approx 0.307$, $\kappa = 2\pi \times 4.29$ MHz and $T_0 = 155$ ns. However, due to the higher order terms in the Hamiltonian, $H_0 = 0$ is realized in the two subspaces with slightly different parameters, $v/2\pi = 98.0$ MHz for $S_z = -1/2$ and

99.8 MHz for $S_z = 1/2$. In the experiment, we use $v/2\pi = 98.8$ MHz to reconcile simultaneous chiral rotations in the two subspaces with $\kappa = 2\pi \times 4.44$ MHz, $f = 2.38$ and $T_0 = 150$ ns.

**Tunable effective coupling between the qubits and the bus resonator** — The periodic modulation of the transition frequencies of the qubits results in effective coupling between the qubits and the bus resonator and between the qubits. In Fig. S1, we show the tunable control of the qubit-resonator coupling. We set the central frequency of the qubit $Q_1$ equal to the resonator frequency and far detune the other four qubits. The effective Hamiltonian between $Q_1$ and the resonator is $H_0 = \hbar g_0 (a^\dagger \sigma_1^- + H.c.)$ with $g_0$ being the effective coupling strength ($g_0 = gJ_0(f)$ in the second order approximation) and $\sigma_j^-$ and $\sigma_j^+$ being the lowing and raising operators. We excite $Q_1$ and measure its probability on the excited state $P_{Q_1}$ as a function of the time delay for different modulating frequencies. We fix the modulating amplitude at $\Delta/2\pi = 235$ MHz. When $v/2\pi \approx 100$ MHz, the effective coupling strength $g_0$ becomes zero and the qubit $Q_1$ is dynamically decoupled from the resonator. From Fig. S1 (b), we can see that the experimental results of decoupling frequency have a deviation about 2 MHz from the zero point of $J_0(f)$, which is due to the high order terms in the effective Hamiltonian.

When we have two qubits coupled with the same resonator, even if the two qubits are dynamically decoupled from the resonator, there is an effective coupling between the two qubits if their modulation phases are different. In Fig. S2, we show the effective coupling between two qubits with modulation frequencies $\omega_j(t) = \omega_0 + \Delta \cos(vt - \phi_j)$. The second order effective interaction Hamiltonian is $H = i\hbar g_{eff} (\sigma_1^+ \sigma_2^- - \sigma_2^+ \sigma_1^-)$ with $g_{eff} = \sum_{n=1}^{\infty} 2g^2 J_n^2(f) \sin(n\Delta\phi)/nv$ where $\Delta\phi = \phi_1 - \phi_2$. We set the parameters such that both qubits are dynamically decoupled from the resonator. We initially prepare $Q_1$ in the excited state and tune $\Delta\phi$. The population swapping between the two qubits are demonstrated by measuring $P_{Q_1}$ as a function of the time delay.

**Numerical simulation** — The numerical simulation with original time-dependent Hamiltonian in Eq. (S1) are done based on QuTiP [1], including the effect from the third level. We use the real experimental parameters of transmons shown in Table S1, where the coupling strength of transmon and resonator is measured on resonance. The coupling strengths and the detunings between the transmons and the cavity suffer some change during the periodic modulation, we use $g' = 1.03g$ and $\delta = \omega_q - \omega_r = -3.5$ MHz with $\omega_q$ being the transition frequency of the qubits in the simulation to fit the experimental data, as shown in Fig. S3. Although the overall initial phase $\phi_0$ has no effect on the effective Hamiltonian, it has a substantial effect on the fast oscillation of the populations [S2]. The value of $\phi_0$ is unknown due to the signal delay in our devices. We adopt $\phi_0 = \pi/3$ to fit the experimental data. The reason of the mismatch between the numerical simulation and the experimental data includes the uncertainties and inhomogeneity in $g'$, $\delta$ and $\phi_0$ for each qubit.

**Any Additional Author notes:** D.-W. Wang conceived the idea and formulated the theory. C. Song performed the experiments. W. Feng, H. Cai and C. Song did the numerical simulation. H. Deng, D. Zheng, and X. Zhu fabricated the sample. D. Xu provided technical support. D.-W. Wang, W. Feng and H. Wang wrote the paper and S. Zhu and M. O. Scully made revisions and comments.